\documentclass[sigconf]{acmart}
\usepackage{hyperref}
\usepackage{url}
\usepackage{graphicx}
\usepackage{etoolbox}

\usepackage[labelformat=simple%,subrefformat=simple
]{subcaption}

\AtBeginDocument{%
  \providecommand\BibTeX{{%
    \normalfont B\kern-0.5em{\scshape i\kern-0.25em b}\kern-0.8em\TeX}}}

\setcopyright{acmcopyright}
\copyrightyear{2020}
\acmYear{2020}
\acmDOI{xx.xxxx/xxxxxxxxx.xxxxxxx}

\acmConference[epiDAMIK 2020]{3rd epiDAMIK ACM SIGKDD International Workshop on Epidemiology meets Data Mining and Knowledge Discovery}{Aug 24, 2020}{San Diego, CA}
\acmBooktitle{epiDAMIK 2020: 3rd epiDAMIK ACM SIGKDD International Workshop on Epidemiology meets Data Mining and Knowledge Discovery}
\acmPrice{15.00}
\acmISBN{978-1-xxxx-XXXX-X}

\makeatletter
\patchcmd{\maketitle}{\@copyrightspace}{}{}{}
\makeatother

\begin{document}

\title{A Deep Learning Approach for COVID-19 Trend Prediction}

\author{Tong Yang}\authornote{Equal Contribution} 
\affiliation{%
	\institution{Department of Physics \\
	Boston College  \\
	Chestnut Hill, Massachusetts, USA}
}

\author{Long Sha}\authornotemark[1]
\affiliation{%
	\institution{Department of Computer Science \\
	Brandeis University  \\
	Waltham, Massachusetts, USA}
}

\author{Justin Li}
\affiliation{%
	\institution{Del Norte High School \\
	San Diego, California, USA}
}

\author{Pengyu Hong}
\affiliation{%
	\institution{Department of Computer Science \\
	Brandeis University  \\
	Waltham, Massachusetts, USA}
}

\begin{abstract}
    The ongoing COVID-19 pandemic, due to the novel coronavirus SARS-CoV-2, has affected not only the healthcare system but the whole society worldwide. 
    While a large number of medical works and researchers are battling the pandemic crisis on the front line, with large amount of accessible epidemic information, data-driven research and learning based approaches could provide rich insights about the challenge on the population and society level.
    In this work, we apply a recurrent-network based model to study the epidemic data in the United States. 
    By incorporating both the epidemic time series and socioeconomic characteristic data, our model provides both a promising predictive power in forecasting the trend of new confirmed cases, and an illustrative description about the interplay between the local epidemic evolution and demographic features.
\end{abstract}

% \begin{abstract}
%     In this work, we developed a deep learning model based approach to forecast the spreading trend of SARS-CoV-2 in the United States. 
%     We implemented the designed model using the Unites States confirm cases and state demographic data, and achieved promising trend prediction results.
%     The model incorporates demographic information and epidemic time-series data through a Gated Recurrent Unit structure.
%     An identification of dominating demographic factors is delivered in the end.
% \end{abstract}

\maketitle

\section{Introduction}
In late 2019, the COVID-19 outbreak initially detected and reported in Wuhan (Hubei, China) due to the \emph{severe acute respiratory syndrome coronavirus 2} (SARS-CoV-2), spread rapidly, firstly across regions in China and east-Asian countries, and then, since late February, to nearly all continents in the world. 
As of \textcolor{black}{June 15}, 2020, there have been more than \textcolor{black}{7.91} million cases confirmed across 225 countries and regions, associated with \textcolor{black}{433} thousands deaths \cite{Yang2020CovidNetTB, 1P3A}. 
Declared as a pandemic by the World Health Organization on March 11, the COVID-19 outbreak has brought severe challenges to not only local healthcare systems (especially in underdeveloped areas) but our society as a whole. 
At the same time, a large number of related research have been emerging recently in various subjects and fields, attempting to contribute to the battle against COVID-19. 
Beside pharmacologic and genomics studies on the SARS-CoV-2 virus, data-driven research both on the spread of COVID-19 among local population and on general social impacts brought by the pandemic have been providing valuable insights especially for local policy makers. 

On the one hand, various types of sequential models have been implemented to study the spreading behavior of COVID-19 in a generic population, including compartmental model based approaches \cite{Cardoso2020UrbanSO, Picchiotti2020COVID19PA, natmed_italy, Dandekar2020.04.03.20052084}, which are motivated by conventional dynamical models in epidemiology, and deep learning based nonparametric approaches \cite{TULI2020100222, Punn2020.04.08.20057679}. 
While the second class, i.e., artificial-learning base approaches, might produce better predictions on disease related statistics, it lacks interpretability for the most part due to the black-box nature of neural network estimators. 

On the other hand, the interaction between social environments and the local COVID-19 outbreak remains an important topic. 
The identification of highly related exogenous factors that impact the local epidemic evolution significantly, e.g., the local population density and the local age structures, is of great importance. 
In addition to understanding dominant environmental factors that govern the outbreak, the relation between the epidemic evolution and socioeconomic characteristics could potentially also reveal the inverse impact of the COVID-19 on a local community \cite{newyork_neighborhood}.

In the current work, we apply a learning-based approach both to produce an accurate prediction about a near future, and to reveal the interplay between environmental factors and the epidemic evolution. 
To accomplish this, we implement a neural-network based sequential model, and integrate the time-varying epidemic information, i.e., related statistics including confirmed cases and deceased records, with environmental factors including both dynamical ones, e.g., local restriction policies, and static ones, such as demographic features.
Environmental factors enter the model via an "kick-start" mechanism~\footnote{See Section~\ref{sec:approach} for details.}, which, after being fixed through training, offers a smoking-gun for the relevance of different factors to the epidemic evolution. 

The rest of the paper is organized as follows. 
Section 2 introduces the ongoing pandemic situation and mentions some related works which motivate our study.
In section 3, we enumerate several candidate factors that could potentially affect the epidemic evolution process significantly and discuss the potential epidemiological dependence, along with other important socioeconomic characteristics, which could be used to analyze the inverse social impact of the ongoing public health crisis.
Section 4 elaborates our application approach in details, explaining both the information flow in the model and the way to extract the relevance of environmental factors to the local epidemic outbreak.
Section 5 explains model training, including data sources, model structures, and training results. 
In section 6, we firstly demonstrate the prediction power of the trained model, and then discuss the relevance of different environmental factors to the epidemic emergence extracted from the trained representation. 
Finally, we summarize our work in Section 7 and discuss potential directions for further investigations with more data and complex models.

\section{The Ongoing Pandemic and Related Works}\label{sec:related}

As of June 15, 2020, the COVID-19 has hit countries worldwide. We summarize the latest pandemic situation in Table~\ref{tab:pandemic}.
\begin{table}[!h]
	\centering
	\begin{tabular}{|c|c|}
		\hline
		\textbf{Category} & \textbf{Statistics} \\
		\hline
		Number of countries reporting COVID-19 cases & 225\\
		\hline
		Number of confirmed cases reported worldwide & 7,912,426\\
		\hline
		Number of deceased cases reported worldwide & 433,391\\
		\hline
		Number of recovered cases reported worldwide & 377,131\\
		\hline
		Currently estimated fatality rate & 5.5\%\\
		\hline
	\end{tabular}
	\caption{Summary statistics of the ongoing COVID-19 pandemic situation worldwide. Data is from the CovidNet project \cite{Yang2020CovidNetTB}.}
	\label{tab:pandemic}
\end{table}

Governments and organizations across the world have been taking measures at different levels in response to this pandemic crisis. 
Extreme measures were adopted by the Chinese government in Wuhan, where a complete lock-down of the whole city was implemented. 
This has been proved later to be very effective to slow down the spread of SARS-CoV-2, the contagious level of which was later revealed to be much higher than two previous deadly viruses, i.e., MERS and SARS.
The response in Wuhan then inspired other countries and regions, including South Korea, Thailand, Italy and etc., addressing the importance of social distancing. 
In spite of the fact that extreme measures have been proven to be effective in fighting against the coronavirus spread, regional lock-down remains a difficult decision to be made for any local government taking into account the economic expense. 
Therefore, it is of extreme importance to provide policy makers necessary tools to both predict the future trend and understand the social impacts brought by the public health emergency \cite{10.1001/jama.2020.5460, social_dist, stay_home}.

On the prediction side, conventional \emph{Susceptible Infectious Recovered}~\footnote{Or \emph{Susceptible Infectious Removed in some literature, which also considered deceased cases.}} ($SIR$) model based approaches have been widely implemented \cite{Cardoso2020UrbanSO, Picchiotti2020COVID19PA, natmed_italy}, with model parameters estimated from regional epidemic data. Motivated by the data-driven estimation process, deep-learning models have also been hybridized into prediction models \cite{Dandekar2020.04.03.20052084}. While the $SIR$ model and its variants indeed could roughly capture the epidemic law of a generic disease spreading behavior, there are two major problems in practice:
\begin{itemize}
    \item ODE systems capture continuous dynamics, howbeit real-life epidemic data is usually collected in discrete time. There are also delays in case reporting, which, even worse, never uniform in time~\footnote{For example, the obvious periodic (week-wise) pattern in the U.S. death data is due to the reporting schedule of the official departments.}. Therefore, there exists a significant mismatch between the true ongoing epidemic process, which can be approximately described by ODEs, and the reported data, which highly depends on human-involved operations.
    \item $SIR$ and its variants only take into account the lowest-order dynamics, which includes the linear terms describing population transitions between compartments and product terms describing interaction/contact between compartments, while keeping transition parameters constants. In reality, however, human responses would also evolve along with the epidemic evolution, which, reversely, could remarkably affect the transmission (due to restriction policies and change in crowd behaviors) and the fatality (due to the improved medical response) of the disease.   
\end{itemize}
Above problems, especially the first one due to human operations, make the task of prediction with compartmental models impractical. 

Beside the predictive power, another drawback of ODE based compartmental models is the absence of environmental factors. 
The trend prediction alone is not enough for designing policy.
Instead, understanding the interplay between environmental factors and the epidemic evolution could benefit local policy makers \cite{10.1001/jama.2020.5460, social_dist, stay_home}, and the dependence of the local outbreak on demographic features and transportation data is essential to calibrate restriction/reopening strategies \cite{Dowd9696}. 
At the same time, it is also of great importance to examine the social impact of the public health emergence on the local community, especially on different population groups characterized by genders, races, and ages.
Most recent works only provide either qualitative arguments \cite{SUN2020e201} or simple statistical analysis, e.g., the linear regression \cite{newyork_neighborhood}, which has limited modeling capability. 

Compared with above methods, our current approach attempts to incorporate environmental factors into the prediction module directly. 
The explicit factor-dependence and therefore interpretability of the model are available via a proper analysis on the learnt representation. 
Details of our modeling and training can be found in Section~\ref{sec:approach} and \ref{sec:training}.

\section{The Interplay between Environmental Factors and the COVID-19 Epidemic}

Before introducing specific model structures and training designs, it is necessary to discuss and distinguish candidate environment factors, which either directly govern the epidemic evolution process, a.k.a. \emph{exogenous factors}, or reveal the social impact of COVID-19 from essential perspectives.

\subsection{Dominant factors of the transmission}

In reality, many exogenous factors would affect the transmission beside disease characteristics. Most intuitively, there is a higher probability in regions with denser population distributions that a fast outbreak would emerge. New York City, being the most densely populated county-equivalence in the United States, would serve as a typical example. 
The outbreak in NYC evolved rather rapidly from the beginning and, as of May 30, 2020, NYC has accounted for than 55 percents of cumulatively confirmed cases across the whole state.

Another dominating factor for transmission would be the restrictive order issued by the local government. 
Restrictions have been implemented onto various industry/business activities as well as local residents' daily life. 
While industry/business restrictions differ region by region, and are usually difficult to study quantitatively, in the present work we instead use the restriction on local residents' daily life, i.e. \emph{the stay-at-home order or its equivalencies}, as an aggregated representation of the \emph{local restriction level} to capture the overall impact of local policies\footnote{Practically, this is also the only consistent data category accessible to the general public}. Importantly, the change of the restrictive level, from the no-restriction stage, to the restrictive-order stage, and finally to the reopen stage, results into a time-varying transmission behavior of COVID-19, which is in contrast to ODE-based compartmental models that assume a constant transmission factor $\beta$.

More generally, the interaction within a local population contributes significantly to the transmission. We therefore adopt the average annual enplanements per capital \cite{enplane} to capture the active level of human interactions.

\subsection{Vulnerable Population Groups and Descriptive Factors}

It has been confirmed by data from multiple countries and regions \cite{doi:10.1056/NEJMoa2002032, 10.1001/jama.2020.4683, cdc_age} that the patient's age is highly related to the development of severe pneumonia symptoms. 
Aged people are in general more vulnerable to the virus. We therefore incorporate the age structure as an important demographic category.

More generally, the physical condition of individuals would result into differences in the probability of infection. 
For instance, people with poor respiratory condition experience higher risk of being infected. 
We therefore include \emph{the population with high risk} \cite{kff_risk} as an input feature of modeling.

\subsection{Smoking-gun factors for social impact analysis}

While above mentioned environmental factors focus more on the biological aspects of the epidemic evolution, there are also other socioeconomic characteristics, which, although not biologically relevant, could be potentially correlated with the evolution behavior. 

For example, a study \cite{newyork_neighborhood} focusing on the New York data has shown that a higher probability of positive testing rate is in poorer neighborhoods, in neighborhoods where large numbers of people reside together, and in neighborhoods with a large black or immigrant population. 
At the same time, however, people residing in poorer or immigrant neighborhoods were less likely to be tested.
As a result, an understanding of which types of neighborhoods are disproportionately affected by the pandemic requires an examination of how socioeconomic characteristics correlate with different epidemic statistics.

Motivated by the above discussion, we selected several socioeconomic characteristics, which are not only related to the epidemic statistics from the pure data perspective, but also important in revealing potential disproportions of the COVID-19 impacts, including local gross domestic product (GDP) per capita and local race compositions.

\begin{figure*}[t]
		\centering
		\begin{minipage}[b]{83mm}
			\centering
			\includegraphics[width=83mm]{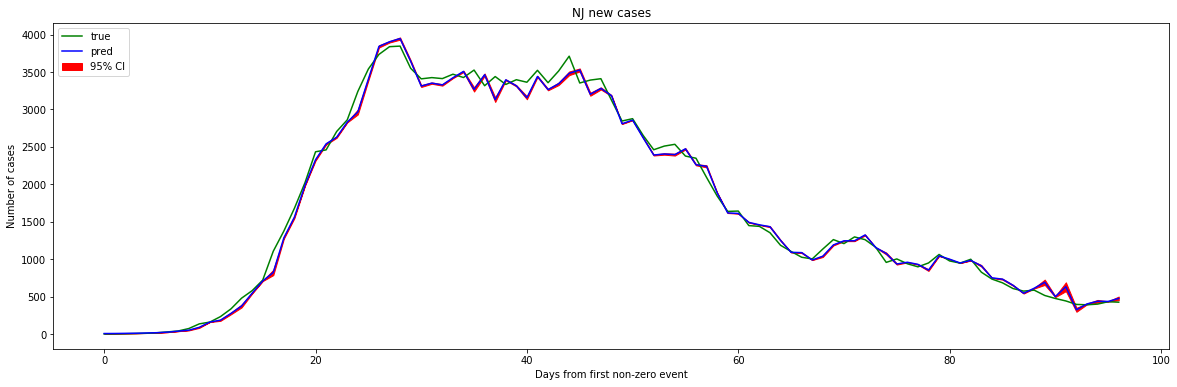}
			\vspace{3.5mm}
			\includegraphics[width=83mm]{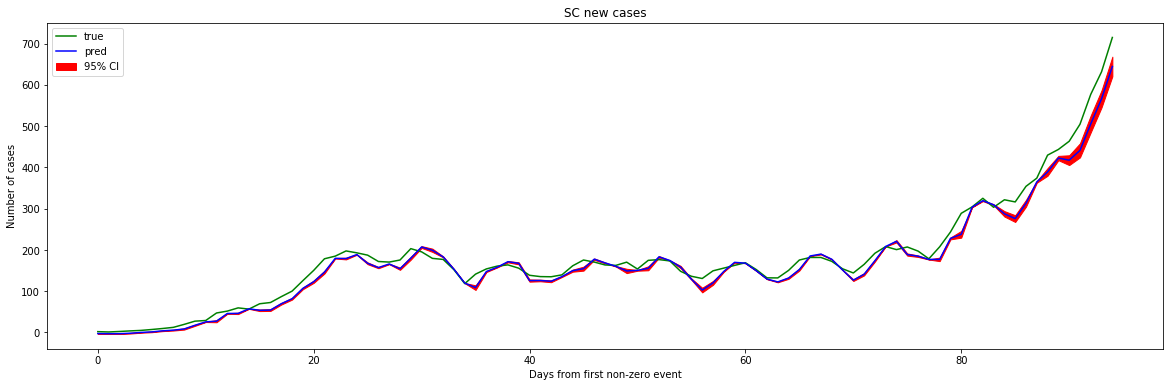}
		\end{minipage}
		\begin{minipage}[b]{83mm}
			\centering
			\includegraphics[width=83mm]{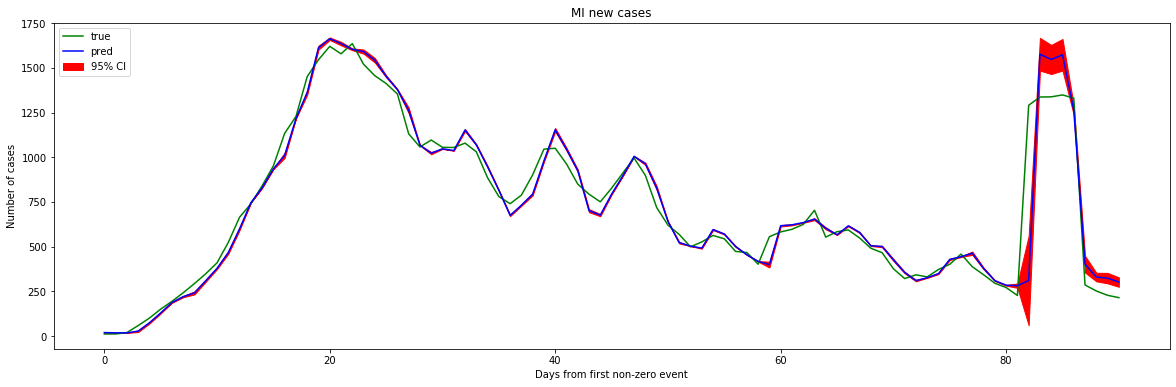}
          \vspace{3.5mm}
          \includegraphics[width=83mm]{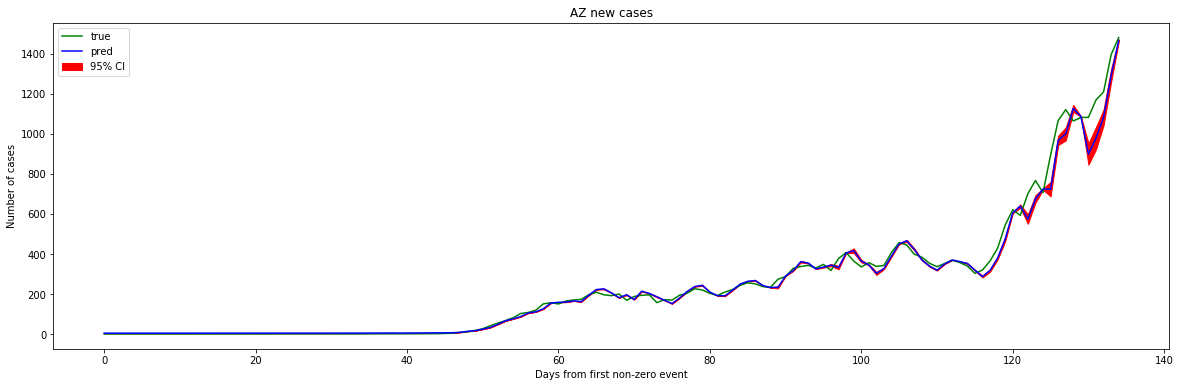}
		\end{minipage}
		
		\vspace{-2mm}
		\caption{\label{fig:new_states}The \emph{1-step-ahead prediction} task on testing states: NJ, MI, SC, AZ, the data from which has never been accessible to the model. The predicted curves follow the true values closely. The 95\% confidence interval regions are obtained by bootstrap re-sampling from the training set.}
		%\vspace{-2mm}
\end{figure*}

\section{A Learning Based Approach}\label{sec:approach}

Now we introduce our learning based approach which, in a nutshell, implements a recurrent-neural-network model for the trend prediction along with an embedding of environmental factors to extract relevant information. 
There are two classes of inputs: the epidemic time series, including both confirmed cases and deceased cases, and environmental socioeconomic factors.

The epidemic time series data enters the learning module through a stacked Gated Recurrent Unit (GRU) model, which is well-known for both its power in dealing with sequential data by incorporating history information properly, and its efficiently simplified structure. 
We cast a fixed-length ($L$) sequence with recent history information to predict each subsequent data point, by implementing a sliding window on the full time series. 
Importantly, we would like to address that this GRU-based model structure is powerful in the following sense:
\begin{enumerate}
    \item Firstly, this GRU model structure, at least, is capable of capturing the dynamics of compartmental models in the discrete-time regime of compartmental models, where the value of the next time-step only relies on the current state. For instance, this can be easily shown through the following set of \emph{difference equations}:
    \begin{align}\label{eq:SIR}
        s_{t+1} - s_{t} &= - \beta\cdot s_{t}\cdot i_{t}, \nonumber \\
        i_{t+1} - i_{t} &= \beta\cdot s_{t}\cdot i_{t} - \gamma\cdot i_{t}, \nonumber \\
        r_{t+1} - r_{t} &= \gamma\cdot i_{t},
    \end{align}
    where $\{s_t, i_t, r_t\}$ represent susceptible, infectious, and removed population fractions respectively, and $\{\beta, \gamma\}$ describe transmission rate and removal rate of the disease. 
    This set of equations can be viewed as a discrete version of a $SIR$ model \cite{Brauer2008}, and they clearly only depend on the current snapshot of epidemic statistics; 
    The above dynamics, in the ideal scenario, can be captured by a GRU model with only $L=1$ sliding-window length.
    \item Mathematically, the above difference equations in Eqs.\ref{eq:SIR} are not always a legitimate format. Indeed, a rigorous transformation from the continuous time regime to the discrete one requires much more caution, and the exact solution form would bring in more complicated terms with both longer history dependence and higher-order polynomial terms. 
    To confront this difficulty, we apply both a sliding window with longer length $L>1$, and more layers in the GRU module to capture complicated nonlinear terms.
\end{enumerate}
Another input feature with time-varying values is the status of the local restriction policy, i.e., the stay-at-home policy or similar measures~\footnote{This includes the stay-at-home advisory issued in Massachusetts, the curfew issued in Puerto Rico, and so on}, which, different from numeric data, is categorical with a binary status: "stay-at-home" or "reopen".
It is naturally expected that the epidemic evolution would be different under different restriction statuses. 
Therefore, we apply a "double-channel" structure in the GRU module: a sequential data point would enter channel-1 if there is a restriction policy on the corresponding date, and would enter channel-2 otherwise.

In addition to time-series data, we have also integrated the following list of environmental factors as static input features:
\begin{itemize}
	\item local population density;
	\item local GDP per capita;
	\item local age structure (fractions of 6 non-overlapping age groups);
	\item local race structure (fraction of 7 different race categories);
	\item high risk population;
	\item local annual enplanements per capital;
	\item local restrictive order level;
\end{itemize}
where all factors are summarized and represented on state level. 
Different from the sequential data of epidemic statistics which enters the model via a black-box, although reasonable, as explained above, model structure, we would like to investigate the interplay between the epidemic evolution and various socioeconomic characteristics. 
Therefore, when we incorporate the input of environmental factors, we apply a linear embedding, which produces interpretable weights on each input dimension.
Technically, the embedded representation of environmental factors is taken as \emph{the initialization of the hidden state in the GRU model}, which we call as a "kick-start" mechanism.

Through the above design of the information flow, we implicitly construct a desired interaction between environmental factors and the epidemic time series data: these two input-categories are conducted to interact with each other via various gates in the GRU module. 
From an epidemic perspective, within the GRU structure, the hidden state could be regarded as an evolving "environment", whose initial status, i.e., before the first infectious case emerges, only depends on exogenous demographic factors of the local community.

\section{Model Training}\label{sec:training}

In this section, we elaborate the practice of model implementation, including data sources, hyperparameters of implemented model, and details of the training procedure. 

\subsubsection*{\textbf{Sources of Different Data Categories}}
\begin{itemize}
    \item \textbf{COVID-19 case data:} Case data is from the CovidNet project \cite{Yang2020CovidNetTB}, including confirmed and deceased counts of 50 U.S. states and the District of Columbia, ranging from January 21, 2020, to June 14, 2020.
    \item \textbf{State restriction policy:} Restriction policy information is collected from "The Coronavirus Outbreak" forum on the New York Times \cite{nytimes}.
    \item \textbf{Population and density:} We have used population data from the U.S. Census Bureau \cite{census}.
    \item \textbf{Population with higher risk:} Population in each state with higher risk to develop severe symptoms are estimated in \cite{kff_risk}, and used as an exogenous factor in our application.
    \item \textbf{Age structure data:} We have used the age structure dataset built by the Kaiser Family Foundation \cite{kff_age}.
    \item \textbf{Race structure data:} Race structure data is collected from the COVID Tracking Project \cite{tracking}.
    \item \textbf{Annual enplanements data:} We collected the data of annual enplanements per capital in each states (not including D.C.) from the U.S. Department of Transportation \cite{enplane}.
    \item \textbf{Gross domestic product per capita:} We collect the data from the United States Census Bureau \cite{census}
\end{itemize}

\subsubsection*{\textbf{Hyper-parameters of the Implemented Model}}

Our model is implemented with an embedding module, recurrent module and output module. The embedding module sparsely encodes the 21-dimension state-specific demographic vector into a 100-dimension vector; the recurrent module is using 3 stacked GRU layers with 100-dimension hidden states, the recurrent module takes the previous embedding result as its first latent state and the windowed state total confirmed cases and new cases as inputs; a dense layer is used for the output layer in order to predict the target. We trained the model using Adam optimizer \cite{kingma2014adam} with 1e-4 learning rate and discounted the learning rate with a factor of 0.3 if the training loss didn't decrease over 20 epochs. The model is trained on an MacBook Pro with 6-Core CPU.

\subsubsection*{\textbf{Details on the Training Design and Process}}

\begin{figure}
    \centering
    \includegraphics[width=0.5\textwidth]{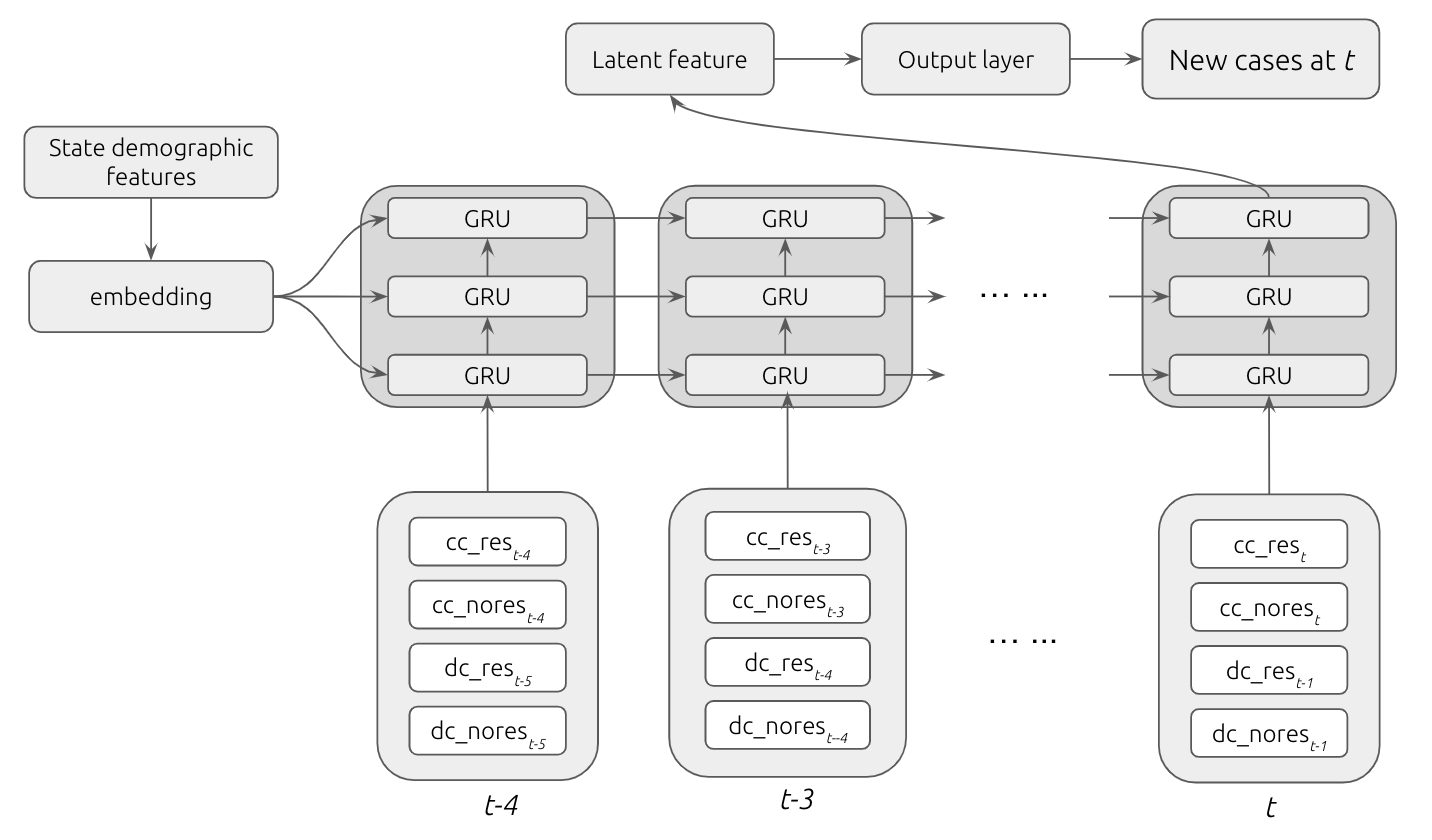}
    \caption{Model architecture}
    \label{fig:flowchart}
\end{figure}

The detailed model architecture is demonstrated in Figure. \ref{fig:flowchart}. We use a five-day window time-series historical data and use recurrent module for prediction. The input data for recurrent module are total confirmed cases ($cc$) and new confirmed cases ($dc$), we pro-process each of them into two series: one contains value only when there is restriction policy undergoing ($cc\_res$ and $dc\_res$) and another only contains value when there is no restriction policy ($cc\_
nores$ and $dc\_nores$) as shown in the model architecture. The processed state demographic data feeds into the state demographic embedding layer, and we use Sigmoid activation function inside the layer. All three layers of GRU receive the embedding output as its first hidden state $h_0$. Root mean squared error(RMSE) is chosen as the loss function. We separate our data firstly withholding 5 states as test data. The others are processed as windowed input-output pairs and separated into a proportion of 80\% for training and 20\% for evaluation. The model is learned via back-propagation utill convergence.

\begin{figure*}[ht]
		\centering
		\begin{minipage}[b]{83mm}
			\centering
			\includegraphics[width=83mm]{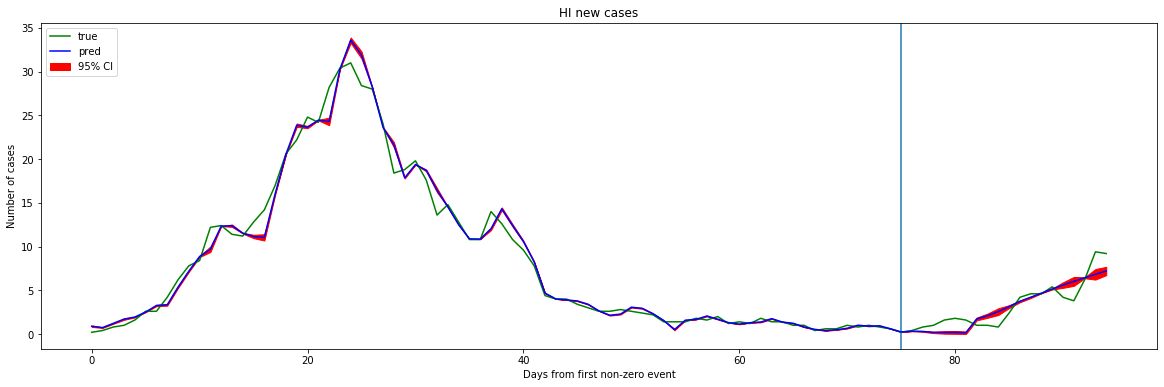}\subcaption{\label{fig:hi}Long-term prediction of Hawaii data.}
			\vspace{3.5mm}
			\includegraphics[width=83mm]{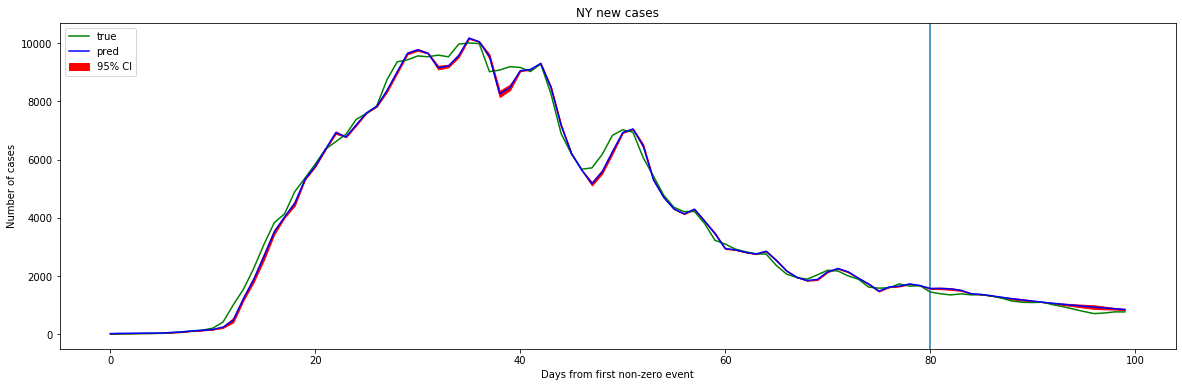}\subcaption{\label{fig:ny}Long-term prediction of New York data.}
		\end{minipage}
		\begin{minipage}[b]{83mm}
			\centering
			\includegraphics[width=83mm]{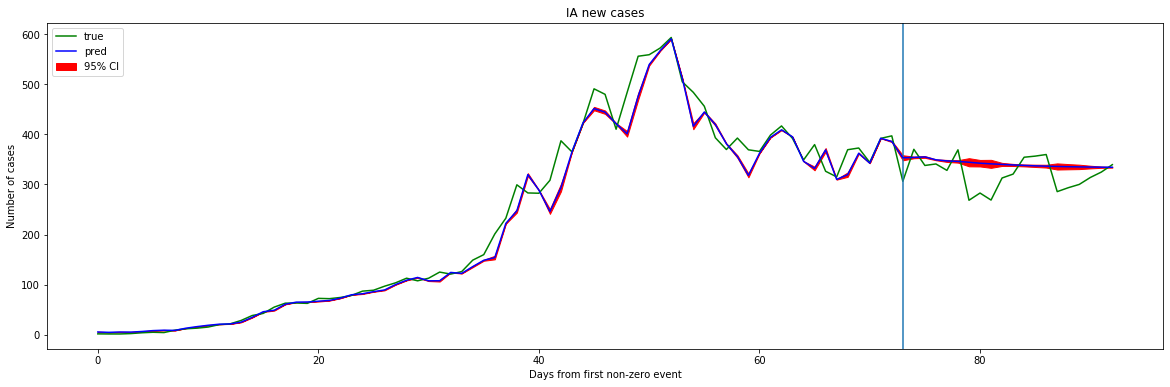}\subcaption{\label{fig:ia}Long-term prediction of Iowa data.}
            \vspace{3.5mm}
            \includegraphics[width=83mm]{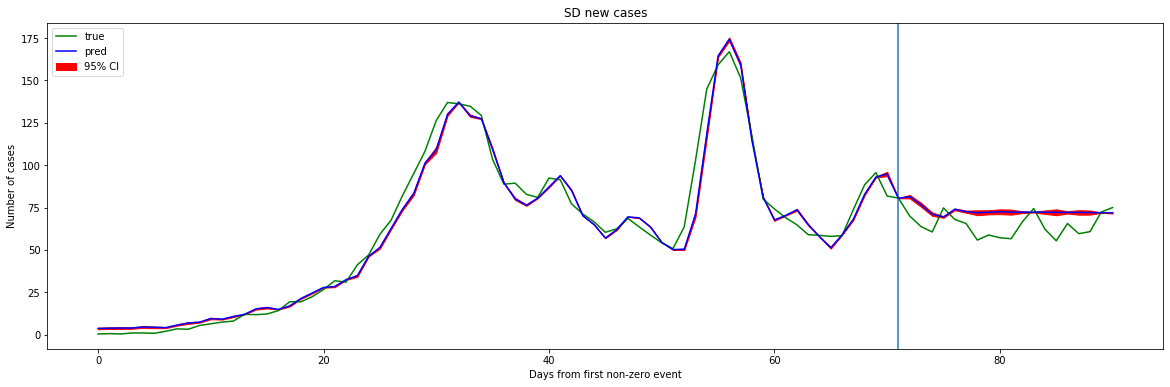}\subcaption{\label{fig:sd}Long-term prediction of South Dakota.}
		\end{minipage}
		
		\vspace{-2mm}
		\caption{\label{fig:ar_pred}The \emph{long-term prediction} task with 4 instantiating states. The solid vertical blue line represents the starting point of the auto-regressive running. The 95\% confidence interval regions are obtained by bootstrap re-sampling from the training set.}
		%\vspace{-2mm}
\end{figure*}
\section{Result Analysis}\label{sec:analysis}

As mentioned earlier, the current work targets both a prediction of the epidemic evolution, and an understanding of the interplay between environmental factors and the local epidemic outbreak. In this section, we discuss the two aspects with the trained model.

\subsection{Prediction of the Epidemic Evolution}

As we have applied the random shuffling during training\footnote{See details explained in Section~\ref{sec:training}} among the training data, which is transformed into \emph{sequence-to-point} pairs, we would demonstrate the prediction power in two ways: 
\begin{enumerate}
    \item "1-step-ahead prediction" on testing states: during the training stage, we have randomly eliminated several states from the complete dataset~\footnote{In our practice, we have randomly selected 4 states for testing purpose: AZ, MI,NJ, SC.}. We would test the performance of the trained model on these states, whose history records have never been acknowledged by the model;
    \item Long-term prediction from an auto-regressive process: the model was trained for 1-step-ahead prediction only during the training stage, therefore long term prediction would be a non-trivial demonstration for the model's capability of capturing the true dynamics.
\end{enumerate}

Figure~\ref{fig:new_states} shows the performance of the model on the \emph{1-step-ahead prediction task}. 
Clearly, even though data from testing states have never been accessible to the model, the trained model can still predict the future value very well.
It is therefore reasonable to state that, rather than over-fitting the given data during the training stage, the model instead capture the general law of the epidemic evolution in a generic population. 
The existence of such a law is not a surprise, and has already been hypothesised in conventional compartmental-model methods. 
However, the general law could easily become intractable from real data due to the human operation in reporting schemes~\footnote{See Section~\ref{sec:related} for detailed discussions.}.
By applying the proposed learning-based method, we extract this general law from the noisy real-life data.

Compared with the \emph{1-step-ahead prediction} task, the \emph{long-term prediction} is much more challenging, in the sense that the task nature has deviated from the training stage. 
The performance of long-term predictions is shown in Figure~\ref{fig:ar_pred}.
While in some states, the deviation from the true data become visible, the overall trend has still been well captured by the auto-regressive process, except noisy fluctuations. 
In practice, the long-term prediction could provide more timely information to policy makers, and hence is more valuable than \emph{1-step-ahead predictions}.

\subsection{Relevance of Environmental Factors}

To reveal the interplay between environmental factors and the epidemic evolution, we start from an analysis on the relevance of each input feature to the dynamics.
As introduced in Section~\ref{sec:approach}, static environmental factors, after being embedded through a linear transformation, enter the model via the "kick-start" mechanism. 
Due to the simple structure of this embedding module, we could easily identify the relevance of input features by examining the Frobenius norm of each embedding vector.

\begin{figure}[ht]
		\centering
		\begin{minipage}[b]{42mm}
			\centering
			\includegraphics[width=42mm]{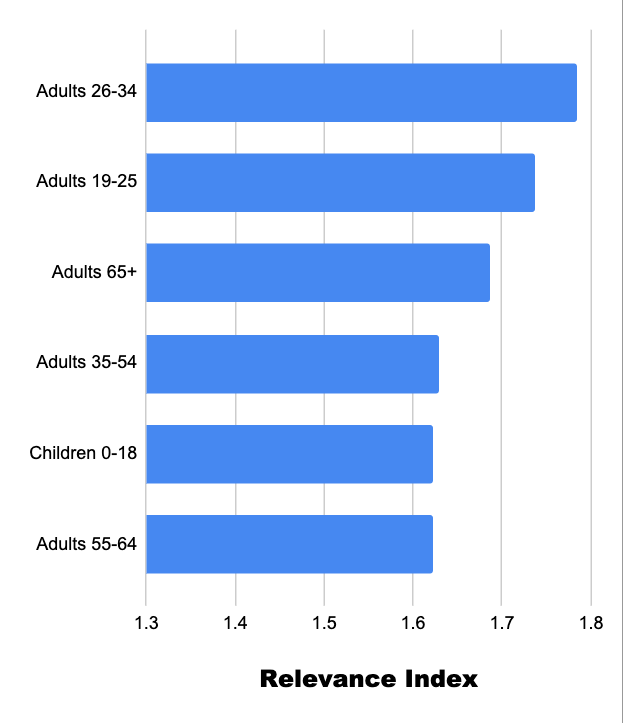}\subcaption{\label{fig:age} Age-group \emph{relevance}.}
		\end{minipage}
		\begin{minipage}[b]{42mm}
			\centering
			\includegraphics[width=42mm]{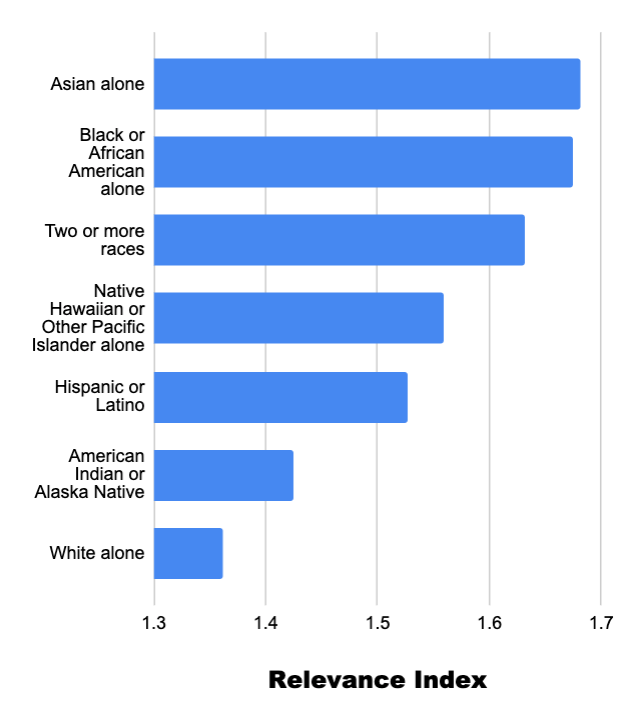}\subcaption{\label{fig:race} Race-group \emph{relevance}.}
		\end{minipage}
		
		\vspace{-2mm}
		\caption{\label{fig:demo_struct} Relevance of different age-group fractions~\ref{fig:age} and race-group fractions~\ref{fig:race}.
		The \emph{relevance index} is defined as the Frobenius norm of the embedding vector of each input feature.}
\end{figure}

Firstly, there are two classes of population structure data: the age structure and the race structure. 
Figure \ref{fig:age} and \ref{fig:race} show the relevance of different age groups and race groups respectively.

In the age group relevance chart, it is clear that the two young-age groups, i.e., \emph{age from 19 to 25} and \emph{26 to 34}, show the highest relevance. 
This is consistent with the demographic report released by CDC on age distribution~\cite{cdc_dist}, where the age group $18-44$ contributes the largest portion to the confirmed cases. 

Among all race groups, the two groups, \emph{Asians} and \emph{Black or African American alone}, appear to be more relevant in epidemic dynamics, while both \emph{White (non-Hispanic)} and \emph{Hispanic/Latino} show lower relevance. 
On the other hand, it is interesting to note that, according to the report released by CDC~\cite{cdc_dist}, \emph{White (non-Hispanic)} and \emph{Hispanic/Latino} contribute largest portion in the confirmed cases.
While the share in confirmed cases may be more related to the absolute population size of different race groups, our relevance analysis, instead, focuses on the fraction of each race group in a certain state.
The above mismatch between the results obtained via the two descriptive perspectives suggests a potentially existing disproportional impacts of the COVID-19 on different groups. 
While the above argument does not provide a rigorous causal analysis, it illustrates the importance of diversity of perspectives when studying the social impact of COVID-19.

Beside the above two types of population structures, we also notice a high relevance ($1.6810614$) of enplanements data to the epidemic dynamics. This confirms our earlier hypothesis that the enplanements data could be used as a nice indicator for the active level of local socioeconomic activities.

\section{Conclusions and Discussions}\label{sec:conclusion}

We have demonstrated the predictive power of the proposed recurrent-network based model, and discussed the relevance of different environmental factors by studying the embedding vector of each socioeconomic characteristic. 

On the prediction side, the proposed model performs well in both \emph{1-step-ahead prediction on new states} and \emph{long-term prediction} tasks. 
One could conclude that the recurrent structure has successfully extracted and captured a general law of the epidemic evolution in a generic population, from the real-life noisy data.

On the other hand, studying the relevance of environmental factors to the epidemic dynamics enables us both to identify potential factors that contribute most to the disease spreading, and to understand the social impact of COVID-19 on the local community.
More specifically, we noticed that young age groups and average emplacements are highly relevant to the dynamics, verifying the fact that socioeconomic activities contribute significantly to the disease spread;
besides, there might exist a disproportion of the social impact on different race groups brought by the COVID-19.

In general, one could expect that more insights about the ongoing public health crisis could be gained through data-driven research. 
Besides medical and clinical studies that directly battle the COVID-19 emergence, it is also important to obtain a more complete understanding about general social impacts of the pandemic on the population and society level. 
This does not only assist local policy makers in decision making, but also helps the whole society to confront the challenge together.

\section{Acknowledgements}
Funding for the shared GPU-computing facility used in this research was provided by NSF OAC 1920147.

\bibliographystyle{ACM-Reference-Format}
\bibliography{main}

\end{document}